\documentclass[aps,11pt, superscriptaddress, showpacs,showkeys,floatfix]{revtex4}

\usepackage{graphicx}

\input{epsf}

\begin{document}
\title{Dynamics of  a liquid dielectric attracted by a cylindrical capacitor}

\author{Rafael Nardi\footnote{E-mail: rafael-uff2003@if.uff.br} and 
 Nivaldo A. Lemos\footnote{E-mail: nivaldo@if.uff.br}}

\address{Departamento de F\'{\i}sica, Universidade Federal Fluminense,\\
Av. Litor\^anea s/n, Boa Viagem  \\
CEP 24210-340, Niter\'oi - RJ, Brazil.}

\date{\today}

\begin{abstract}
The dynamics of a liquid dielectric attracted by a vertical cylindrical capacitor is studied. Contrary to what might be expected from the standard 
calculation of the force 
exerted by the capacitor, the motion of the dielectric is different 
depending on whether the charge or the voltage of the capacitor is held constant. The problem turns out to be an unconventional example 
of dynamics of a system with variable mass, whose   velocity can, in certain circumstances, suffer abrupt changes.
Under the hypothesis that the voltage remains constant the motion is described in qualitative and quantitative details, and a very brief  qualitative 
discussion is made of the constant charge case.

\end{abstract}

\pacs{41.20.Cv, 45.20.D-, 45.20.da}

\keywords{cylindrical capacitor, force on a dielectric, variable-mass system}

\maketitle

\section{Introduction}
Electric forces  on insulating bodies are notoriously difficult to calculate \cite{Stratton}.
Only in very special instances  can the calculation  be performed by  elementary means. The simplest case,
treated in every undergraduate textbook on electricity and magnetism,
 is that of a  dielectric slab attracted by a parallel-plate capacitor.
Nearly as simple is the determination of the force of attraction on a liquid
dielectric by a vertical cylindrical capacitor. It is remarkable that, under appropriate conditions,  in
both cases
 conservation of energy alone  leads to a simple expression for the force
 with no  need to know the complicated fringing field
responsible
for the attraction. Conservation of energy leads to the same expression for
the force independently of whether the voltage or the charge of the capacitor
is assumed to remain constant during a hypothetical infinitesimal  displacement of the dielectric.
It is reassuring that a detailed treatment based on a  direct consideration
of the fringing field leads to the same result \cite{Dietz}.
Since  the  dynamics of the dielectric under such
a force is not discussed in the textbooks, one may be left with the false impression that the motion 
is unaffected by the choice of constant  voltage or constant  charge.
   
  In this paper we study the dynamics of a liquid dielectric pulled
into a  vertical cylindrical capacitor by electrostatic forces. The motion is quantitatively different,
though qualitatively the same, depending on whether the voltage or the charge
of the capacitor is held constant. It will be seen that this  is a very peculiar
variable-mass system, since in certain circumstances the 
 velocity can change abruptly during a very short time interval right after the beginning of the motion.

     The paper is organized as follows. In Section 2 we review the elementary
calculation of the force exerted by a vertical cylindrical capacitor on a
liquid dielectric. In Section 3 the dynamics is investigated assuming the
voltage applied to the capacitor stays constant. The motion of the liquid near the bottom edge is described qualitatively only,
since our expression for the electrostatic force is not valid near the ends of the capacitor. For the motion when the surface of the dielectric is 
well inside the capacitor quantitative  results are obtained.
  In Section 3 we briefly  discuss the case in which  the charge of the capacitor
is held fixed, and a qualitative description of the motion is
given.   Section 4 is devoted to final comments and conclusions.

\section{Force of a cylindrical capacitor on a liquid dielectric}

Let us consider a coaxial cylindrical capacitor of length $L$ whose inner and outer radii are $a$ and $b$, respectively, with $L \gg b$. Suppose the 
capacitor is in the vertical position and is partially immersed in a liquid dielectric
contained  in a tank, in such a way that  the surface of the liquid that has entered the  capacitor is far from its ends.  If a potential difference is applied between the 
capacitor plates, the dielectric will
be pulled into the capacitor. The calculation of  the force of attraction is a textbook problem \cite{Griffiths} which we proceed to work out  for subsequent 
use.

\begin{figure}[h]
\epsfxsize=7cm
\begin{center}
\leavevmode
\epsffile{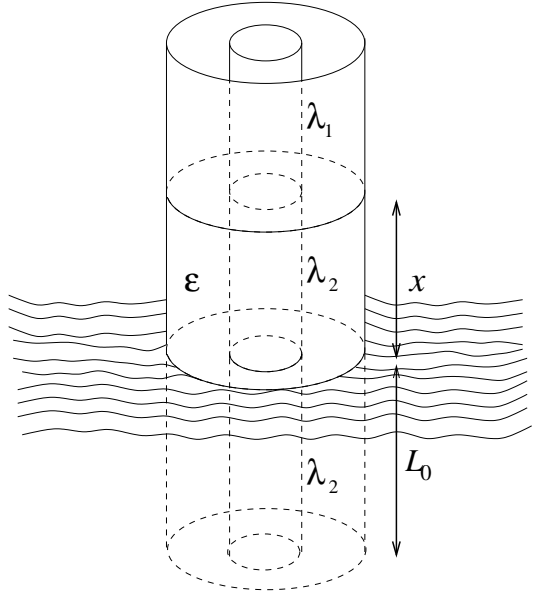}
\caption{A cylindrical capacitor partially immersed in a tank containing a liquid dielectric. The liquid inside  the capacitor rises when a  potential difference is applied 
between the plates.}
\end{center}
\end{figure}

 Let  $ L_0$ be the length of the capacitor that is immersed in the liquid before  the voltage is turned on. After the voltage is turned on
 the liquid inside the capacitor rises to the height  $ L_0 +x$ ,  as depicted in Fig 1. Let $\lambda_1$ and $\lambda_2$ be the linear charge 
densities 
on the parts of the 
inner plate  that are in empty space and in contact with the dielectric, respectively. The corresponding electric fields in vacuum and inside the
 dielectric are

 \begin{equation}
\label{electricfields}
{\bf E}_1= \frac{\lambda_1}{2\pi\epsilon_0 s}\,{\hat{\bf s}}\, ,\,\,\, {\bf E}_2= \frac{\lambda_2}{2\pi\epsilon s}\,{\hat{\bf s}}
\, ,
\end{equation}
where $ {\hat{\bf s}}$ is the radial unit vector and $\epsilon$ is the permittivity of the dielectric. The voltage $V$ applied to the capacitor is
 given by the radial line 
integral either of ${\bf E}_1$ or ${\bf E}_2$:

 \begin{equation}
\label{voltage}
V= \int_a^b {\bf E}_1 \cdot {\hat{\bf s}}\, ds=\frac{\lambda_1}{2\pi\epsilon_0}\ln\frac{b}{a}
=  \int_a^b {\bf E}_2  \cdot {\hat{\bf s}}\, ds=\frac{\lambda_2}{2\pi\epsilon}\ln\frac{b}{a}
\, .
\end{equation}
From this the capacitance can be readily found. Indeed, the total charge of the capacitor is 

 \begin{equation}
\label{charge}
Q = \lambda_1 [L - (L_0 + x)] +  \lambda_2  (L_0 + x) = \frac{2\pi V}{\ln (b/a)}\bigl\{\epsilon_0  [L - (L_0 + x)] + \epsilon (L_0 + x) \bigr \}
\, ,
\end{equation}
whence  

 \begin{equation}
\label{capacitance}
C  = \frac{2\pi \epsilon_0}{\ln (b/a)}[L + \chi_e( L_0 +  x)]
\, ,
\end{equation}
where $ \chi_e$ is the electric susceptibility of the dielectric.

The force on the dielectric can be determined by equating the work done by the capacitor during a displacement $dx$ of the dielectric 
to the  decrease of the electrostatic 
energy stored in the capacitor. Thus, assuming the charge $Q$ of the capacitor remains constant as the displacement takes place,

  \begin{equation}
\label{force}
F = - \frac{dE}{dx} =  - \frac{d}{dx}\bigg(\frac{Q^2}{2C}\bigg) =  \frac{1}{2}\frac{Q^2}{C^2}\frac{dC}{dx}=
 \frac{\pi\epsilon_0 \chi_e V^2}{\ln (b/a)}
\, ,
\end{equation}
where Eq. (\ref{capacitance}) has been used. Since this force is positive, the dielectric is pulled into the capacitor.
As is well known, the same result is obtained for the 
force if, instead of the charge  $Q$, the voltage $V$ of the capacitor is assumed to remain constant during the displacement.
But in the latter case the battery responsible for  maintaining the voltage  fixed does work as the dielectric is displaced, and this work
 must be taken into account in the energy balance \cite{Griffiths}. Since the {\it expression} for the force is the same in both cases, it may seem obvious
 that the {\it motion} of the liquid dielectric does not 
depend on whether the charge or the voltage is held constant. This is not the case, however, for a  reason that may be easily overlooked.  
If the voltage is held constant
then the force (\ref{force}) is constant during the entire motion. However,  if the charge $Q$ remains constant then, using $V=Q/C$ and
 Eq. (\ref{capacitance}),
 the force  (\ref{force}) becomes

 \begin{equation}
\label{force-constantQ}
F =   \frac{Q^2}{C^2}
 \frac{\pi\epsilon_0 \chi_e }{\ln (b/a)} =  \frac{Q^2}{4\pi\epsilon_0}\frac{ \chi_e \ln (b/a)}{[L + \chi_e (L_0 +  x)]^2}
\, .
\end{equation}
This is  a {\it variable} force that depends on the position of the dielectric because the capacitance increases and the voltage decreases as
the dielectric penetrates the capacitor. Therefore, the {\it dynamics} of the dielectric is different  depending on
whether the charge or the voltage of the capacitor is  constant throughout the  motion.

\section{Equation of motion of the liquid dielectric}

The equation of motion of the liquid dielectric is

\begin{equation}
\label{equationmotion}
\frac{d}{dt} (m{\dot x}) = F + \rho g L_0 \pi (b^2 -a^2) - mg
 \end{equation}
where $m$ is  the mass of the dielectric that is already inside the capacitor at time $t$. The presence of the term  proportional to $L_0$ on the right-hand side
of  (\ref{equationmotion}) is justified as follows. The pressure at the bottom of the liquid column that is inside the 
capacitor is higher than the pressure at the top by  $\rho g L_0$, where $\rho$ is the density of the dielectric. This gives rise to a net upward pressure
 force that equals $ \rho g L_0 \pi (b^2 -a^2)$ and must be added to the electrostatic force $F$.
We have  assumed 
 that the area of the tank is
 much larger
 than $\pi(b^2-a^2)$, so that the variation of the height of the liquid in the tank can be safely ignored.  Other hypotheses will be mentioned in 
due time.


 The mass of liquid inside the capacitor is
 
\begin{equation}
\label{mass}
m = \pi (b^2 -a^2)(L_0 + x)\rho\, .
 \end{equation}
Insertion of  this expression  for the mass into the equation of motion (\ref{equationmotion}) leads to

\begin{equation}
\label{equationmotionsimplyfied}
\frac{d}{dt} \bigl[ (L_0 + x){\dot x}\bigr] = \frac{F}{ \pi (b^2 -a^2) \rho} - g x
\, .
 \end{equation}

\section{Motion in the constant voltage case}

If $V$ is constant, the use of (\ref{force}) allows us to write Eq.(\ref{equationmotionsimplyfied}) as

\begin{equation}
\label{equationmotion-constantV-new variable-y}
y{\ddot y}+ {\dot y}^2 + gy=gL_0 +  \frac{\epsilon_0 \chi_e V^2}{\rho  (b^2 -a^2) \ln (b/a) }
\, .
 \end{equation}
where we have introduced the new variable $y$ defined by 

\begin{equation}
\label{newvariable-y}
y= L_0 + x
\, .
 \end{equation}

The task of solving equation (\ref{equationmotion-constantV-new variable-y}) is made easier by writing it in dimensionless form. It is easy to check that the positive constants 
$a_0$ and $t_0$ defined by

\begin{equation}
\label{scales-lenght-time-V}
a_0= \frac{\epsilon_0 \chi_e V^2}{\rho g  (b^2 -a^2) \ln (b/a) }=gt_0^2
 \end{equation}
have dimensions of length and time, respectively. In terms of the dimensionless variables $\xi$ and $\tau$, and dimensionless parameter $\alpha$, defined by

\begin{equation}
\label{dimensionless-variables}
\xi = \frac{y}{a_0}=\frac{L_0 + x}{a_0} \, ,\,\,\, \tau = \frac{t}{t_0} \, ,\,\,\, \alpha =\frac{L_0}{a_0} \, ,
 \end{equation}
the differential equation (\ref{equationmotion-constantV-new variable-y}) reduces to

\begin{equation}
\label{dimensionless-equationmotion-constantV}
\xi \frac{d^2\xi}{d\tau^2} + \bigg(\frac{d\xi}{d\tau}\bigg)^2 +\xi =1 + \alpha
\, .
 \end{equation}
This equation has the equilibrium solution $\xi_{eq} =1+ \alpha$, which is tantamount to $x_{eq}=a_0$. Thus, $a_0$ is the height at which the weight of the dielectric
that has risen above the external level   is exactly counterbalanced by the electric   force that pulls it up.

Equation  (\ref{dimensionless-equationmotion-constantV}) can be solved by standard techniques. First of all we make a change of independent variable:

\begin{equation}
\label{change-independent-variable}
p= \frac{d\xi}{d\tau} \Longrightarrow  \frac{d^2\xi}{d\tau^2}=   \frac{dp}{d\tau}=  \frac{dp}{d\xi}  \frac{d\xi}{d\tau}= p \frac{dp}{d\xi}
\, .
 \end{equation}
This reduces Eq.(\ref{dimensionless-equationmotion-constantV})  to

\begin{equation}
\label{equationmotion-constantV-for-p}
\xi p \frac{dp}{d\xi} + p^2 + \xi -1-\alpha  =0
\, ,
 \end{equation}
or, in terms of differential forms,

\begin{equation}
\label{equationmotion-constantV-for-p-differential}
 \xi pdp + (p^2 + \xi -1 - \alpha)d\xi =0
\, .
 \end{equation}

We seek an integrating factor $N(\xi )$ that turns the left-hand side of this  equation into an exact differential:

\begin{equation}
\label{integrating-factor}
N(\xi ) \xi pdp + N(\xi ) (p^2 + \xi -1 - \alpha)d\xi =dG
\, .
 \end{equation}
Such a function $G(\xi, p)$ exists if and only if

\begin{equation}
\label{solutionchange-iintegrating-factor}
\frac{\partial}{\partial \xi}[ N(\xi )\xi p ]=\frac{\partial}{\partial p}[ N(\xi ) (p^2 + \xi -1 - \alpha) ]   \Longrightarrow \xi \frac{dN}{d\xi}=N
\, .
 \end{equation}
Thus, $N(\xi )=\xi$ is an integrating factor and we have

\begin{equation}
\label{differential-of-G}
dG= \xi^2 pdp + \xi  (p^2 + \xi -1 - \alpha)d\xi 
\, .
 \end{equation}
The function $G$ can be readily found by integrating $dG$ along the broken path composed
 of the two straight line segments $(0,0)\to (\xi ,0)\to (\xi ,p)$, and one immediately gets

\begin{equation}
\label{solution-for-G}
G(\xi ,p) = \frac{1}{2}\xi^2 p^2 +  \frac{\xi^3}{3} - \frac{ 1+ \alpha}{2}\,\xi^2
\, .
 \end{equation}
Therefore, the general solution of (\ref{equationmotion-constantV-for-p}) is

\begin{equation}
\label{solution-for-p}
 \frac{1}{2}\xi^2 p^2+ \frac{ \xi^3}{3}  - \frac{ 1+ \alpha}{2}\,\xi^2 = C
\, ,
 \end{equation}
where $C$ is an integration constant.

\subsection{Close to the edge: a qualitative description}

Equation (\ref{force}) is adequate for the force only when the surface of the liquid is far from the edges of the capacitor. Therefore, 
it is not legitimate to put  $L_0=0$ in equation (\ref{equationmotion-constantV-new variable-y}).  Nevertheless, something remarkable happens when one 
considers  $L_0=0$. 
Since the attraction is caused by the fringing field, it is clear that  the force on the liquid dielectric is 
not zero when its surface just touches the lower edge of the capacitor. 
Suppose that at $t=0$ the liquid is at rest,  the voltage is turned on  
and the upward motion begins with $\xi =0$. Then the ``first" ascending  layer of the liquid has zero mass but is pulled into the capacitor with a finite force.
Thus, ideally, it is subjected 
to an infinite acceleration, which brings about a finite jump of the velocity. Sure enough  there is a transient process during which the electric field builds up
 and the 
velocity changes continuosly. However, since the time scale of this transient process  is certainly  much shorter than the  time scale
of the motion, it seems reasonable to expect that it can be satisfactorily
modelled by
a discontinuous jump of the velocity.  Furthermore, the value attained by the velocity right after the transient, when so to speak 
 the electrostatic force takes over, 
most likely cannot be freely chosen. This is a very peculiar behavior of this unconventional variable-mass system, whose mass can drop to zero at  
certain instants in the course of its motion.

It turns out that this unique  behavior  is
nicely  illustrated, at least in some  
of its main qualitative aspects, by our description. In order to check this statement let us try to find out  what the equation of motion has to tell us about the dynamics of the 
liquid near the edge of the capacitor, although the quantitative results are admittedly unreliable.

With $\alpha =0$ because we are taking  $L_0=0$, the 
 initial condition $\xi (0) =0$ demands
$p(0) =\pm 1$, as plainly seen  from either (\ref{dimensionless-equationmotion-constantV}) or (\ref{equationmotion-constantV-for-p}). Accordingly, 
since $\xi$ starts to increase immediately after $t=0$, we take for
 initial conditions $\xi (0) =0$ and $p(0) = 1$, a choice that leads to  $C=0$ in equation (\ref{solution-for-p}). As a consequence, 
 Eq.(\ref{solution-for-p}) can be solved for $p$ in the form

\begin{equation}
\label{solution-for-tau}
 \frac{d\xi}{d\tau} =\sqrt{1-\frac{2}{3}\xi}\,\,\, \Longrightarrow \,\,\, 
\tau =\int_0^{\xi} \frac{d\lambda}{\sqrt{1-\frac{2}{3}\lambda}}= -3\sqrt{1-\frac{2}{3}\xi} + 3
\, .
 \end{equation}
This is  easily inverted to yield

\begin{equation}
\label{solution-for-xi}
\xi (\tau )= \frac{9- (\tau -3)^2}{6} =  \frac{\tau (6 - \tau)}{6}
\, .
 \end{equation}

It is clear that $\xi$ attains its maximum value $\xi_{max}=3/2$ at $\tau =3$, and returns to its initial value $\xi =0$ at $\tau =6$. Then $p$ changes 
abruptly from $p=-1$ to $p=1$ and the motion repeats itself, as shown in Fig. 2. The period of oscillation is $\tau_0 =6$ in dimensionless units. 
These quantitative results are not trustworthy and the cusps at the turning points will be smoothed out in the actual motion, but the qualitative nature of
 the dynamics  seems to be well represented by our description.


\begin{figure}[h]
\epsfxsize=9cm
\begin{center}
\leavevmode
\epsffile{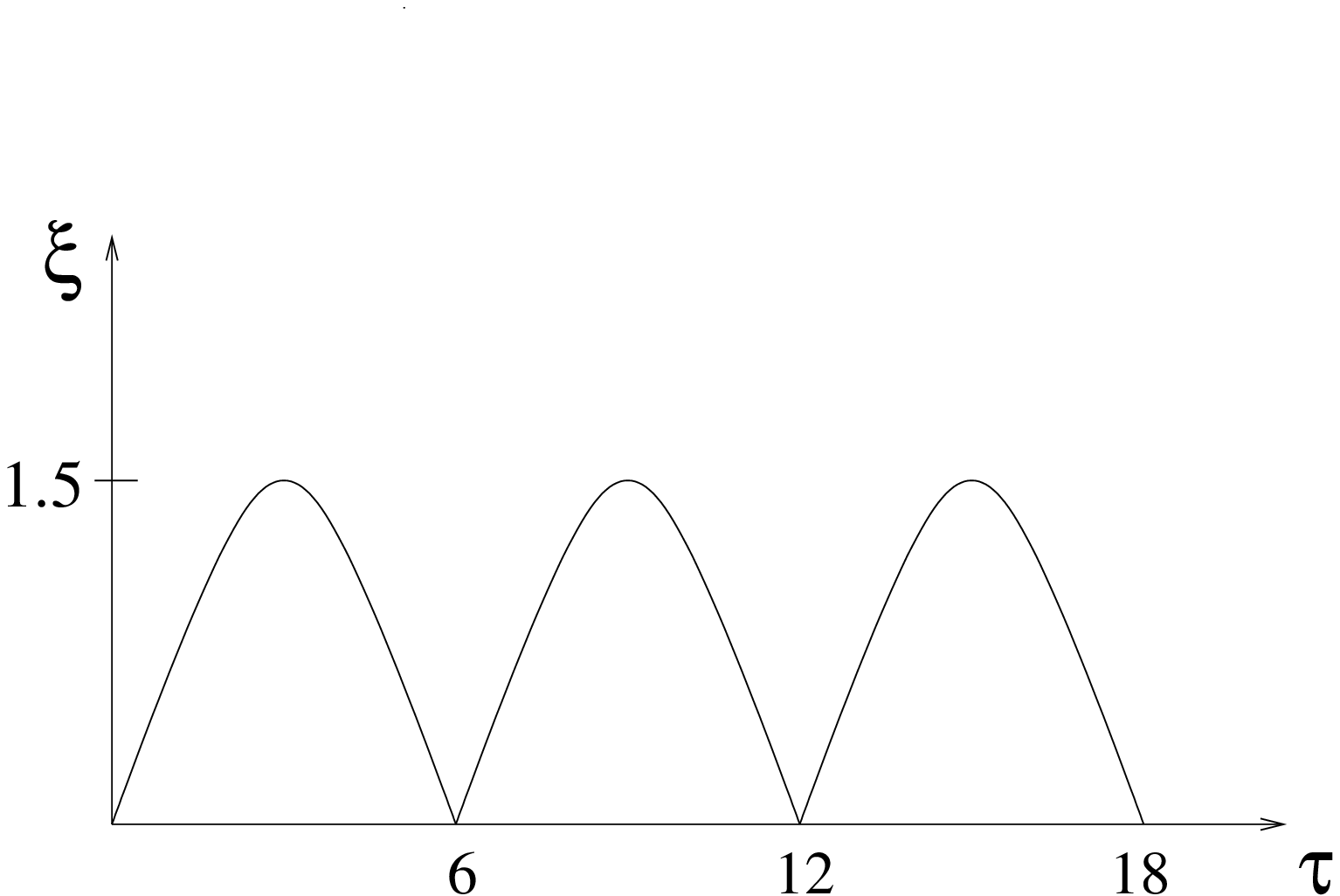}
\caption{Periodic oscillation of the surface of the liquid dielectric near the bottom edge of the capacitor 
in terms of dimensionless quantities in the constant voltage case.}
\end{center}
\end{figure}

\subsection{Far from the edge: a quantitative description}

Let us suppose  now that   $L_0 \approx L/2$, where $L$ is the 
length of the capacitor,  so that   the liquid starts its motion well inside the capacitor. In this case our appoach is expected to be
fairly realistic,  and furnish  quantitative results that can be taken seriously. If the fluid is at rest when the voltage is turned on,
the initial conditions are  $x(0)=0$ and ${\dot x}(0)=0$, which  are equivalent to $\xi (0)=\alpha$ and ${\dot \xi}(0)=0$ 
according to equation (\ref{dimensionless-variables}). In this case the constant of 
integration is given by $\,C=\alpha^3/3 - (1+\alpha)\alpha^2/2$ and  equation
(\ref{solution-for-p}) becomes

\begin{equation}
\label{equation-for-p-far-from-edge}
 \frac{1}{2}\xi^2 p^2+ V(\xi )=0\, ,\,\,\,\ V(\xi )= \frac{ \xi^3 -\alpha^3 }{3}  - \frac{ 1+ \alpha}{2}\,(\xi^2 - \alpha^2)
\, .
 \end{equation}

Recalling that $p=d\xi/ d\tau$, this equation can be solved in the form

\begin{equation}
\label{solution-for-csi-far-from-edge}
\tau = \frac{1}{\sqrt{2}}\int_{\alpha}^{\xi}  \frac{z \, dz}{\sqrt{\frac{1+\alpha}{2}(z^2 -\alpha^2) -  \frac{1}{3}(z^3-\alpha^3)   }}
\, .
 \end{equation}
The appearence of the square root of a cubic  polynomial in the above integrand indicates that the solution for $\xi(\tau )$ will be given by elliptic functions. 
However, much 
interesting information about the motion can be obtained without using these not so familiar functions. To this end note that 
equation  (\ref{equation-for-p-far-from-edge})  is formally equivalent to the one that describes the motion of a particle with 
zero ``total energy" in the ``potential" $V(\xi )$. The positivity of the ``kinetic energy" $ \xi^{2} {\dot \xi}^2/2$ constrains $\xi$ to the region $V(\xi )\le 0$. 
Since $x\ge 0$, in terms of $\xi$  the physically allowed region is  $\xi \ge \alpha$, as follows from Eq.(\ref{dimensionless-variables}).

 The function $V(\xi )$ has the following properties: (i) $V(\alpha )=0$; (ii) $V^{\prime}(\alpha )=-\alpha <0$ which implies
$V(\xi)<0$
for $\xi$ slightly bigger than $\alpha$; (iii) $V(\xi)\to\infty$ as $\xi\to\infty$; (iv) there is only one positive solution to $V^{\prime}(\xi )=0$, which 
 is readily found to be $\xi_{eq}= 1+ \alpha$ and
 corresponds to $x_{eq}=a_0$.  This result for  $\xi_{eq}$ is merely a confirmation of what was found directly from the equation of
 motion (\ref{dimensionless-equationmotion-constantV}).
 Now we have enough information  to draw a qualitative graph of the ``potential", which is depicted in Fig. 3. The motion is confined to the 
``potential well"  shown
in the same figure.
\begin{figure}[h]
\epsfxsize=8cm
\begin{center}
\leavevmode
\epsffile{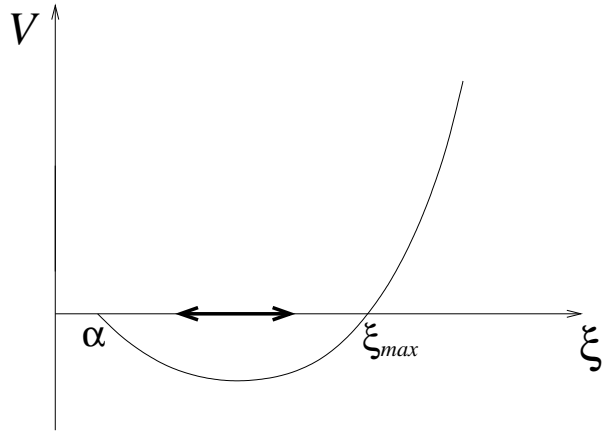}
\caption{Qualitative description of the periodic  oscillation of the surface of the liquid dielectric  in the constant voltage case. The 
dimensioless variable $\xi$ is confined in the  ``potential well"   between $\xi =\alpha$ and $\xi=\xi_{max}$.}
\end{center}
\end{figure}


The amplitude of the oscillation of the surface of the liquid, that is, the maximum value attained by
$\xi$ is the only positive value  $\xi=\xi_{max}>\alpha$ such that $V(\xi )$ equals zero. In the region $\xi >\xi_{max}$ the ``potential" is positive and the   
motion is forbidden. The amplitude $\xi_{max}$ is the positive solution  to  the algebraic equation

\begin{equation}
\label{roots-potential}
 \frac{\xi^3 - \alpha^3}{3}= \frac{1+ \alpha}{2}(\xi^2 -\alpha^2)\,\, \Longrightarrow\,\,  \frac{\xi^2 + \alpha\xi +\alpha^2}{3}= 
 \frac{1+ \alpha}{2}(\xi +\alpha )
\, ,
 \end{equation}
since $\xi\neq \alpha$. This quadratic equation for $\xi$ is solved by

\begin{equation}
\label{equation-for-csi-max}
\xi_{max}=\frac{1}{4}\bigg[ 3+ \alpha + 3\alpha \, \sqrt{1+ \frac{10}{3\alpha}+ \frac{1}{\alpha^2}}\, \bigg] 
\, .
 \end{equation}
Typically, as will be seen presently, $L_0 \gg a_0$, so that $\alpha \gg 1$. In this case a 
 binomial expansion  gives the following  approximate expression for  $\xi_{max}$:

\begin{equation}
\label{approximate-equation-for-csi-max}
\xi_{max}=\frac{1}{4}\bigg[  \alpha + 3 + 3\alpha \, \Bigl(1+ \frac{5}{3\alpha}\Bigl)\, \bigg] = \alpha + 2
\, .
 \end{equation}
As a consequence, $x_{max} = 2 a_0$. Therefore the surface of  the liquid dielectric inside the capacitor oscillates periodically 
 between $x=0$ and $x=x_{max}=2a_0$. From (\ref{solution-for-csi-far-from-edge})  it follows at once that the period of  oscillation is

\begin{equation}
\label{period-far-from-edge}
T = \sqrt{2}\, t_0\,\int_{\alpha}^{\xi_{max}}  \frac{\xi \, d\xi}{\sqrt{\frac{1+\alpha}{2}(\xi^2 -\alpha^2) -  \frac{1}{3}(\xi^3-\alpha^3)   }}
\, .
 \end{equation}
An expression for $T$ in terms of elliptic integrals is given in the Appendix.

Unfortunately, under normal laboratory conditions the oscillation of the surface of the liquid will be very hard to observe. In order to make
 $a_0$  and $t_0$ as large as possible, one needs a low-density liquid with very high susceptibility.  Not only $a$ and $b$ but also their difference
should be as small as possible, while 
the voltage should be as high as possible. However, these choices are limited, among other things,
by the dielectric strength of the liquid.  The  dielectric strength of typical insulators \cite{Forsythe} is about $10^7$ V/m. The maximum electric field
 inside the capacitor is 
easily seen to be $E_{max}= V/ a \ln (b/a)$, so that one must have $V< 10^7 \,a \ln (b/a)$, with $a$ and $b$ in meters. Furthermore, 
the difference $b-a$ cannot be too small lest capillarity becomes dominant. To give an idea of the order
 of magnitudes involved, consider 
pure water, for which  $\rho \approx 1\, \mbox{g/cm}^3$ and $\chi_e \approx 79$. Take  $a=1.0\, \mbox{cm}$, $b=2.0\, \mbox{cm}$,
 $L=20\, \mbox{cm}$,  $L_0=10\, \mbox{cm}$, and  
$V=5000$ volts, a very high voltage. Then from (\ref{scales-lenght-time-V}) and (\ref{dimensionless-variables}) we obtain
$ a_0=0.86\,$cm, $t_0=30\,\mbox{ms}$  and $\alpha =11.6$, so that $\xi_{max}=13.6$. It follows that  the
 amplitude of the oscillation  is $x_{max}\approx 2 a_0 = 1.7\,\mbox{cm}$. The 
 period of oscillation calculated in the Appendix is $T= 0.93\,\mbox{s}$. It should be stressed that these values for the amplitude and
 the period require an
 extremely high voltage.

\section{Constant charge case}

Making use of (\ref{force-constantQ}) the equation of motion (\ref{equationmotionsimplyfied}) becomes

\begin{equation}
\label{equationmotion-constantQ}
\frac{d}{dt} [ (L_0 +x){\dot x}]= \frac{\chi_eQ^2\ln (b/a)}{4\pi^2 \epsilon_0 \rho (b^2-a^2)}\frac{1}{(L + \chi_{e}x)^2} -gx
\, .
\end{equation}
The same  change of variables (\ref{dimensionless-variables}),  but this time with
$a_0$ and $t_0$ replaced by

\begin{equation}
\label{scales-lenght-time-Q}
{\bar a}_0 = \frac{\chi_eQ^2\ln (b/a)}{4\pi^2 \epsilon_0 \rho (b^2-a^2)L^2 g}
\,\, \,  , \, \, \, {\bar  t}_{0} =\sqrt{ \frac{\chi_eQ^2\ln (b/a)}{4\pi^2 \epsilon_0 \rho (b^2-a^2)L^2g^2}}
\, ,
\end{equation}
reduces equation (\ref{equationmotion-constantQ}) 
to the dimensionless form

\begin{equation}
\label{dimensionless-equationmotion-constantQ}
\xi \frac{d^{2}\xi}{d{\tau}^{2}} + \left(\frac{d\xi}{d\tau}\right) ^{2} + \xi = \alpha +  \frac{1}{(1+ \sigma \xi)^{2}},
\end{equation}
where $\alpha = L_0/{\bar a}_0$ and the additional  dimensionless parameter $\sigma$ is given by

\begin{equation}
\label{dimensionless-parameter}
\sigma = \frac{\chi_{e}{\bar a}_0}{L}
\, .
\end{equation}

The same procedure as in the previous case leads to the following first integral for Eq. (\ref{dimensionless-equationmotion-constantQ}):

\begin{equation}
\label{first-integral-Q}
\frac{\xi^{2}p^{2}}{2} + \frac{\xi^{3}}{3} - \frac{\alpha}{2}\xi^2  - \frac{1}{\sigma^{2}} \left[\ln (1 + \sigma \xi) + \frac{1}{1 + \sigma \xi} \right] 
= {\bar C}
\, . 
\end{equation}
The initial conditions $\xi(0) = \alpha$,  $p(0)=0$ determine   ${\bar C}$.
A qualitative description of the motion can be given following the pattern set by the previous case. The shape of the ``potential" is exactly the same as the one
 shown in Fig. 3. The motion is periodic and confined to the interval
 $\alpha \le \xi \le {\bar \xi}_{max}$ where, now, ${\bar \xi}_{max}$ is determined by a transcendental  equation because of the logarithmic term in 
(\ref{first-integral-Q}), which prevents us from giving an explicit expression for  the amplitude of the oscillation. The period is also given by a complicated
 integral involving 
the square root of a logarithm.

\section{Conclusion}

The motion of a liquid dielectric attracted by a  vertical cylindrical capacitor is an unusual  example of dynamics of a variable-mass system.
As we have seen, because the initial mass vanishes for motion of the liquid near the lower edge of the capacitor with
 the most natural initial conditions, the velocity change can be qualitatively
modelled by a  jump
determined  by the equation of motion itself.  Assuming that the surface of the liquid remains far from the ends of the capacitor, we
 exactly solved for the motion in the case of constant voltage and found not too complicated 
 expressions for the amplitude and  the period of the oscillation.  If, instead, the charge of the capacitor is held constant, the motion is qualitatively the
 same but quantitative 
results are different, and demand more complicated numerical computations.

\begin{acknowledgments}
The work of Rafael Nardi was partially supported by Conselho Nacional de Desenvolvimento Cient\'{\i}fico e Tecnol\'ogico
(CNPq), Brazil.
\end{acknowledgments}

\appendix

\section{The period in terms of elliptic integrals}

According to (\ref{roots-potential}) the ``potential" $V(\xi )$ has the three  real roots $\xi_1=\alpha$, $\xi_2=\xi_{max}\equiv \beta$, 
$\xi_3= - \gamma$, where

\begin{equation}
\label{realroots-V}
\beta=\frac{1}{4}\bigg[ \alpha +3 + 3\alpha\sqrt{1+ \frac{10}{3\alpha} + \frac{1}{\alpha^2}} \, \bigg]
\approx \alpha + 2\, , \,\, \gamma =\frac{1}{4}\bigg[ 3\alpha\sqrt{1+ \frac{10}{3\alpha}
 + \frac{1}{\alpha^2}}  -\alpha -3 \bigg]
\approx \frac{1+\alpha}{2}
\end{equation}
and the approximate equalities hold if $\alpha \gg 1$. Thus we can write 

\begin{equation}
\label{fractorization-V}
 -V(\xi )= \frac{ 1+ \alpha}{2}\,(\xi^2 - \alpha^2) - \frac{ \xi^3 -\alpha^3 }{3}   = \frac{1}{3}(\xi - \alpha)(\beta -\xi)(\xi + \gamma)
\end{equation}
and equation  (\ref{period-far-from-edge})  for the period takes the form

\begin{equation}
\label{integral-period-Appendix}
T = \sqrt{2}\, t_0 \int_{\alpha}^{\beta} \frac{x\, dx}{\sqrt{-V(x)}} = \sqrt{6}\, t_0 \,
 \int_{\alpha}^{\beta} \frac{x\, dx}{\sqrt{(x-\alpha)(\beta - x) (\gamma + x)}}\equiv \sqrt{6}\, t_0 \, I \, .
\end{equation}
We follow  \cite{Spiegel}  and perform the change of variable $x = \beta - u^2$,   which leads to

\begin{equation}
\label{integral-period-variable-u}
I= 2\int_{0}^{\sqrt{\beta - \alpha}} \frac{(\beta - u^2)du}{\sqrt{(\beta-\alpha - u^2)(\beta  + \gamma  - u^2)}} \equiv 2\beta I_1 - 2I_2\, .
\end{equation}
The additional change of variable $u=\sqrt{\beta - \alpha}\,\sin\theta$ reduces both integrals $I_1$ and $I_2$ to the standard form of complete elliptic integrals. Let us first tackle $I_1$:

\begin{equation}
\label{first-integral-theta}
I_1= \int_{0}^{\sqrt{\beta - \alpha}} \frac{du}{\sqrt{(\beta-\alpha - u^2)(\beta +\gamma  - u^2)}}=
 \frac{1}{\sqrt{\beta + \gamma}}\int_{0}^{\pi/2} \frac{d\theta}{\sqrt{1- k^2\sin^2\theta}}\,,
\end{equation}
that is,

\begin{equation}
\label{first-integral-elliptic}
I_1=     \frac{1}{\sqrt{\beta + \gamma}}\, K(k)\, ,\,\,\,      k=\sqrt{\frac{\beta -\alpha}{\beta + \gamma}}
\end{equation}
where $K(k)$ is the  complete elliptic integral of the  first  kind. Similarly

\begin{equation}
\label{second-integral-theta}
I_2 =     \frac{\beta - \alpha}{\sqrt{\beta + \gamma}}\int_{0}^{\pi/2} \frac{\sin^2 \theta\, d\theta}{\sqrt{1- k^2\sin^2\theta}} = 
\frac{\beta - \alpha}{\sqrt{\beta + \gamma}}\int_{0}^{\pi/2} \frac{-(1-k^2\sin^2 \theta )/k^2 + 1/k^2 }{\sqrt{1- k^2\sin^2\theta}} d\theta\, ,
\end{equation}
so that

\begin{equation}
\label{second-integral-elliptic}
I_2 =     \frac{\beta - \alpha}{k^2\sqrt{\beta + \gamma}}\bigg[ \int_{0}^{\pi/2} \frac{d\theta}{\sqrt{1- k^2\sin^2\theta}} -
 \int_{0}^{\pi/2} \sqrt{1- k^2\sin^2\theta}\, d\theta\bigg] = 
 \sqrt{\beta + \gamma}\, \Bigl[ K(k) - E(k)\Bigr] \, ,
\end{equation}
where $E(k)$ is the complete elliptic integral of the  second kind.

After some simplifications, we finally get

\begin{equation}
\label{period-elliptic-integrals}
T = 2\sqrt{6}\, t_0\, \bigg\{\sqrt{\beta +\gamma}\, E(k) - \frac{\gamma}{\beta +\gamma} K(k)\bigg\}\, .
\end{equation}
The functions $K(k)$ and $E(k)$ are given by the following power series ($0\leq \vert k\vert  < 1$):

\begin{equation}
\label{series-elliptic-integral-K}
K(k) =  \int_{0}^{\pi/2} \frac{d\theta}{\sqrt{1- k^2\sin^2\theta}} = \frac{\pi}{2}\bigg\{ 1  + \bigg(\frac{1}{2}\bigg)^2 k^2  + \bigg(\frac{1\cdot 3}{2\cdot 4}\bigg)^2 k^4 + \bigg(\frac{1\cdot 3\cdot 5}{2\cdot 4 \cdot 6}\bigg)^2 k^6 + 
\ldots \bigg\}\, ;
\end{equation}

\begin{equation}
\label{series-elliptic-integral-E}
E(k) = \int_{0}^{\pi/2} \sqrt{1- k^2\sin^2\theta}\, d\theta =\frac{\pi}{2}\bigg\{ 1  - \bigg(\frac{1}{2}\bigg)^2 k^2  - \bigg(\frac{1\cdot 3}
{2\cdot 4}\bigg)^2 \, \frac{k^4}{3} - \bigg(\frac{1\cdot 3\cdot 5}{2\cdot 4 \cdot 6}\bigg)^2 \, \frac{k^6}{5} + \ldots \bigg\}\, .
\end{equation}

For the parameters chosen in the main text we have seen that  $\alpha =11.6$. Then (\ref{realroots-V})  yields $\beta =13.6$ and $\gamma =6.26$. With these values the
modulus of the complete elliptic integrals is found from (\ref{first-integral-elliptic}) to be $k=0.317$. For this $k$ the values for the complete
 elliptic integrals furnished by the above power series are $K(k)=1.6125$ and $E(k)=1.5306$. Finally, with   $t_0 = 30\, \mbox{ms}$ equation 
(\ref{period-elliptic-integrals}) gives $T = 0.93\, \mbox{s}$. 


\end{document}